\documentclass[%
 reprint,
 amsmath,amssymb,
 aps,
]{revtex4-2}
\long\def\comment#1{}

\usepackage{graphicx}
\usepackage[latin1]{inputenc}
\usepackage{amsfonts}
\usepackage{amsmath,amssymb}
\usepackage{bm}
\usepackage{ulem}
\usepackage{color}
 \long\def\comment#1{}

\begin{document}
\graphicspath{{images/}}

\title{Charging capacitors using diodes at different temperatures. II Numerical studies}
\author{J.M. Mangum,$^1$ L.L. Bonilla,$^2$ A. Torrente,$^2$ and P.M. Thibado,$^{1,*}$}
\affiliation{$^1$Department of Physics, University of Arkansas, Fayetteville, Arkansas 72701, USA.\\
$^2$G. Mill\'an Institute for Fluid Dynamics, Nanoscience and Industrial Mathematics and Department of Mathematics, Universidad Carlos III de Madrid, 28911 Legan\'es, Spain.\\
$^*$Corresponding author. E-mail: thibado@uark.edu}
\date{\today}

\begin{abstract}
This study is presented in a series of two papers. The first paper is an analytical study. This is the second paper, and here we numerically study the thermal energy harvesting capability of two electronic circuits. The first circuit consists of a diode and capacitor in series. We solve the time-dependent Fokker-Planck equation and show the capacitor initially charges and then discharges to zero. The peak charge on the capacitor increases with temperature, capacitance, and diode quality. The second circuit has two current loops with one small capacitor, two storage capacitors, and two diodes wired in opposition. When the diodes are held at different temperatures we observe a non-zero steady-state charge is accumulated on both storage capacitors. The magnitude of the stored charges are nearly equal but the signs are opposite. When resistors are used in place of diodes there is no transient and no steady-state charge buildup. Numerical studies for the time-independent Fokker-Planck equation are presented and confirm the steady state charges.
\end{abstract}

\maketitle
\begin{center}
\textbf{I. INTRODUCTION}
\end{center}

Recent breakthroughs in electronic circuit designs have enabled ultralow levels of power consumption. Some sensor systems, for example, draw only picowatts in standby mode, and can be programmed to rarely be active~\cite{han09}. This technological breakthrough creates the possibility of powering devices by scavenging energy from the local ambient environment.

One important source of local power is thermal. For example, with two different temperatures, one can generate an electrical voltage using the thermoelectric effect~\cite{seebeck}. In these materials, charge carriers diffuse from the hot side to the cold side to create a voltage. Thermoelectric properties are generally enhanced for materials having a high ratio for electrical conductivity and thermal conductivity~\cite{chang,gayner}. Thermoelectric devices are typically made from bismuth telluride~\cite{fernandez,ezzi}. Other materials like tin selenide have been shown to have a comparable thermoelectric efficiency and are less expensive~\cite{zhao}. It has been proposed that quantum well superlattices may yield an improvement in thermoelectric efficiency~\cite{hicks}. Other materials such as carbon nanotubes, silicon nanowires, and graphene have also been shown to operate as a heat engine and to have favorable thermoelectric qualities~\cite{ack16, durbin, blackburn,collins,bouki,dragoman,dollfus}. 

Other options, beyond thermoelectrics, for harvesting energy from temperature gradients are available. For example, rectify thermal fluctuations in the presence of a temperature gradient using nonlinear electrical devices has been proposed~\cite{kam57,kam60}. In the late 1990's, Sokolov theoretically studied systems using capacitors and diodes~\cite{sok98,sok99}. It was found that when there is a temperature difference between the diodes, charge is stored on the capacitor in the steady state. Later, it was found that nonlinear devices can be used to harvest thermal energy even at a single temperature by taking advantage of transients~\cite{thi20, thi23}.

In this study, we present coordinated results for the role nonlinear devices play in thermal energy harvesting across two papers in a series. The first paper presents analytical solutions. In this second paper, we present numerical solutions. We completed the numerical study for two circuits. The first has one diode and one capacitor arranged in a single loop. The second circuit has two diodes which can be held at different temperatures. The second circuit also has two storage capacitors and two current loops. We present results showing both charging in the transient dynamics and in the steady state.

\maketitle
\begin{center}
\textbf{II. SINGLE LOOP CIRCUIT}
\end{center}

A schematic diagram for the first circuit we investigate is shown in Fig. 1(a). This circuit has a diode and capacitor in series. We also include a DC bias voltage (normally set to zero), which allows us to study its role in the transient dynamics. The Hamiltonian for this circuit is given by 

\begin{eqnarray}
\mathcal{H}=\frac{q^2}{2C}+qV,
\end{eqnarray}
where $q$ is the charge on the capacitor, $C$ is its capacitance, and $V$ is the bias voltage. 
We model the diode conductance ($\mu$) using a sigmoid function given by

\begin{eqnarray}
\mu(u)=\frac{1}{R} \frac{1}{1+e^{-u/u_0}},
\end{eqnarray}
where $u$ is the voltage across the diode [found using Eq. (1)], $R$ is the resistance at high forward bias, and $u_0$ sets the quality of the diode. For example, as $u_0$ approaches zero, the diode conductance matches that of an ideal switch, and as $u_0$ becomes larger the diode conductance becomes a linear resistor. Another benefit of the sigmoid function is that it mimics an ideal diode in series with a resistor, which more closely represents a real diode~\cite{sze81}. To study the circuit dynamics we solve the Fokker-Planck equation given by

\begin{eqnarray}
\frac{\partial \rho}{\partial t}=\frac{\partial}{\partial q}\left( \mu \rho \frac{\partial \mathcal{H}}{\partial q} \right) + k_BT \frac{\partial}{\partial q}\left( \mu \frac{\partial \rho}{\partial q}\right),
\end{eqnarray} 
where $\rho$ is the probability distribution function for the charge on the capacitor at time $t$, $k_B$ is Boltzmann's constant, and $T$ is the absolute temperature.

In this study, we solve the various Fokker-Planck equations using Mathematica. We use the nonlinear partial differential equations solver which can require a lot of computer memory (up to one terabyte). The initial probability distribution function (pdf) is set to be a Gaussian close to a Dirac delta function centered at zero charge. Mathematica uses a variety of numerical methods and automatically switches between nonstiff and stiff methods as needed. It employs the method of lines and a range of solvers including Runge-Kutta, implicit Euler, and extrapolation methods. Once complete Mathematica finds the full pdf in time. We first confirm the area under the probability distribution function stays one throughout the simulation, and then use the pdf to calculate the various moments.

\begin{figure}[ht]
\begin{center}
\includegraphics[width=8cm]{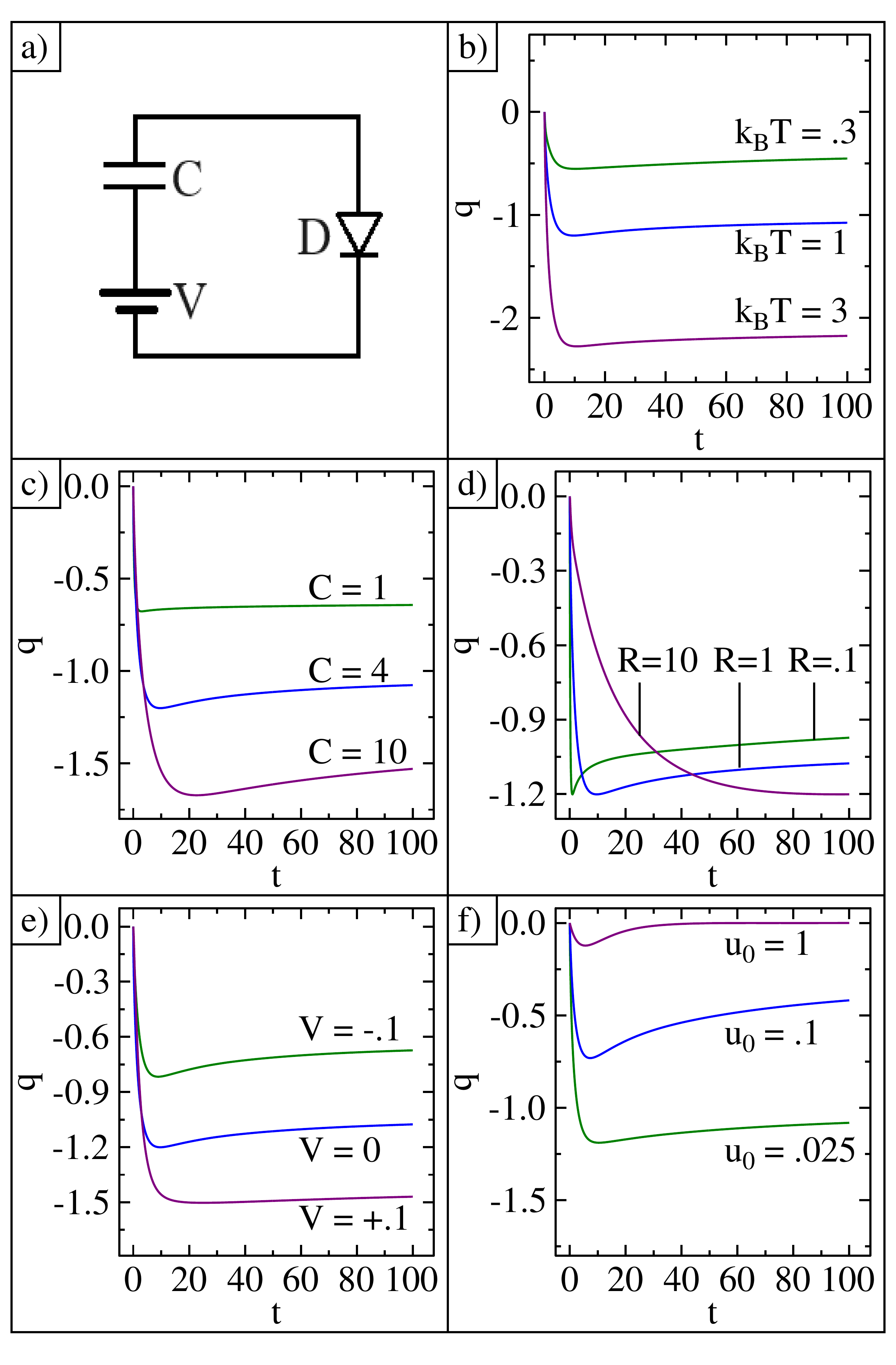}
\end{center}
\caption{Single-loop circuit with a capacitor, diode, and a DC bias voltage.~Numerical results use the following parameters $u_0=0.025$, $R=1$, $C=4$, $k_BT=1$, and $V=0$, unless otherwise stated in the figure. (a)~Schematic of the circuit. Average charge on the capacitor in time is shown with alternative parameter values for (b)~temperature, (c)~capacitance, (d)~resistance, (e)~bias voltage, and (f)~diode parameter.}
\end{figure}

\maketitle
\begin{center}
\textbf{A. Single diode}
\end{center}

The average charge on the capacitor in time is shown for a large set of parameter values in Fig. 1. The parameter values used for the numerical studies are $u_0$=0.025, $R$=1, $C$=4, $k_BT$=1, and $V$=0, unless otherwise noted in the figure. In all cases, the charge on the capacitor is set to zero in the beginning and it rapidly charges to various negative values. After some time a peak charge is reached and the capacitor begins to discharge. The peak charge is higher when $k_BT$ is higher, as shown in Fig. 1(b), and when $C$ is higher, as shown in Fig. 1(c). Notice the time to reach the peak charge increases with $C$. Increasing the value of $R$ also increases the time to reach the peak charge but does not alter its magnitude as shown in Fig.~1(d). Adding a bias voltage does impact the dynamics as shown in Fig.~1(e). The overall charge changes by an amount equal to $CV$ as expected, but the capacitor tends to have a longer transient period at higher positive bias. The role of the diode parameter $u_0$ on the charging dynamics is shown in Fig.~1(f). When the diode is close to an ideal switch, at low $u_0$, the capacitor charges and stays charged for a very long time. As the diode becomes closer to a linear resistor, at low $u_0$, only a small charge is formed before discharging.

Understanding the charging dynamics shown in Fig.~1 ultimately comes from the time evolution of the probability distribution (not shown). The initial charge distribution is a delta function centered at zero charge, while the final charge distribution is a Gaussian with average charge $q = CV$ and variance $k_BTC$~\cite{johnson,nyquist}. The diode conductance play a large role in how the distribution evolves from start to finish. In the case of an ideal switch, the charge can only flow in one direction. As a result, at steady state only half of the final Gaussian distribution is present, while the other half stays at zero probability. Therefore, the ideal switch produces an average charge given by $-\sqrt{2k_BTC/\pi}$~\cite{thi23}. Even if the diode is not an ideal switch, the difference between forward and reverse conductance alters the distribution to be asymmetric, and yields the nonzero transient charge on the capacitor. The series of plots shown in Fig. 1(f) shows how the average charge increases as the value of $u_0$ decreases toward an ideal switch. Given that the width of the equilibrium Gaussian distribution increases as $k_BTC$ increases, the average charge will also increase with $k_BT$ and $C$ as shown in Fig. 1(b) and 1(c). The time scale for Eq. (3) is given by $\tau=RC$, which explains the charging time differences shown in Fig. 1(c) and 1(d). When $V$ is positive, the battery drives current clockwise, like the diode, and therefore increases the average charge by an amount equal to $q=CV$ as shown in Fig. 1(e).

The results shown throughout Fig. 1 capture well the full essence and importance of our discovery. The nonlinearity of the diode, when first connected to a capacitor, actually charges it. This does not happen when the diode is replaced with a resistor. Also, higher quality diodes charge the capacitor to a higher level. Of course, in equilibrium the charge distribution on the capacitor will be a Gaussian with mean zero and variance $k_BTC$, which satisfies the second law. A higher quality diode, due to its large reverse bias resistance, will significantly delay the time to equilibrium. As a result, the charge distribution quickly changes from the initial Dirac delta function to a half-Gaussian, and then it slowly evolves to a full Gaussian. Notice clockwise current has a very short time constant (for positive charge carriers), while counterclockwise current has a very long time constant. The half-Gaussian has an average charge different from zero. A higher temperature diode and a larger capacitance capacitor will increase the maximum charge reached because the variance is linear in temperature and capacitance. From an applications perspective, one might use the nonlinear diode to charge a storage capacitor and then disconnect the storage capacitor before it starts to discharge. In principle, diode nonlinearity represents a new source of power, because the stored charge can be used later to power a device.

\maketitle
\begin{center}
\textbf{B. Multiple diodes in series or parallel}
\end{center}

Given that diodes are essential for a transient charge to form, we next study the role of adding more diodes to the circuit on the stored charge dynamics. To do this we replaced the single diode with a set of identical diodes connected together either in series or parallel, as illustrated in the two circuit schematics shown in Fig. 2(a). Each diode has the same sigmoid conductance as before, and we use the same parameter values as before given by $u_0=0.025, R=1, C=4, k_BT=1$, and $V=0.$ For a series of identical diodes, the equivalent conductance is given by

\begin{eqnarray}
\mu_s(u)=\frac{1}{N_sR} \frac{1}{1+e^{-u/N_su_0}},
\end{eqnarray}
where $N_s$ is the number of diodes in series. The resistance of each diode adds together to increase the total resistance to $N_sR$. Also, and more importantly, notice the voltage drop across each diode is not $u$, but a fraction of this value given by $u/N_s$. Having a lower diode voltage significantly lowers its conductance. The average charge of the capacitor in time for different numbers of diodes in series is shown in Fig. 2(b). The maximum charge significantly decreases as the number of diodes increases. In addition, the time to reach the maximum charge significantly increases with the number of diodes. Both the lower individual diode voltage and the higher overall resistance are responsible for the observed changes to the charging dynamics.

\begin{figure}[ht]
\begin{center}
\includegraphics[width=8cm]{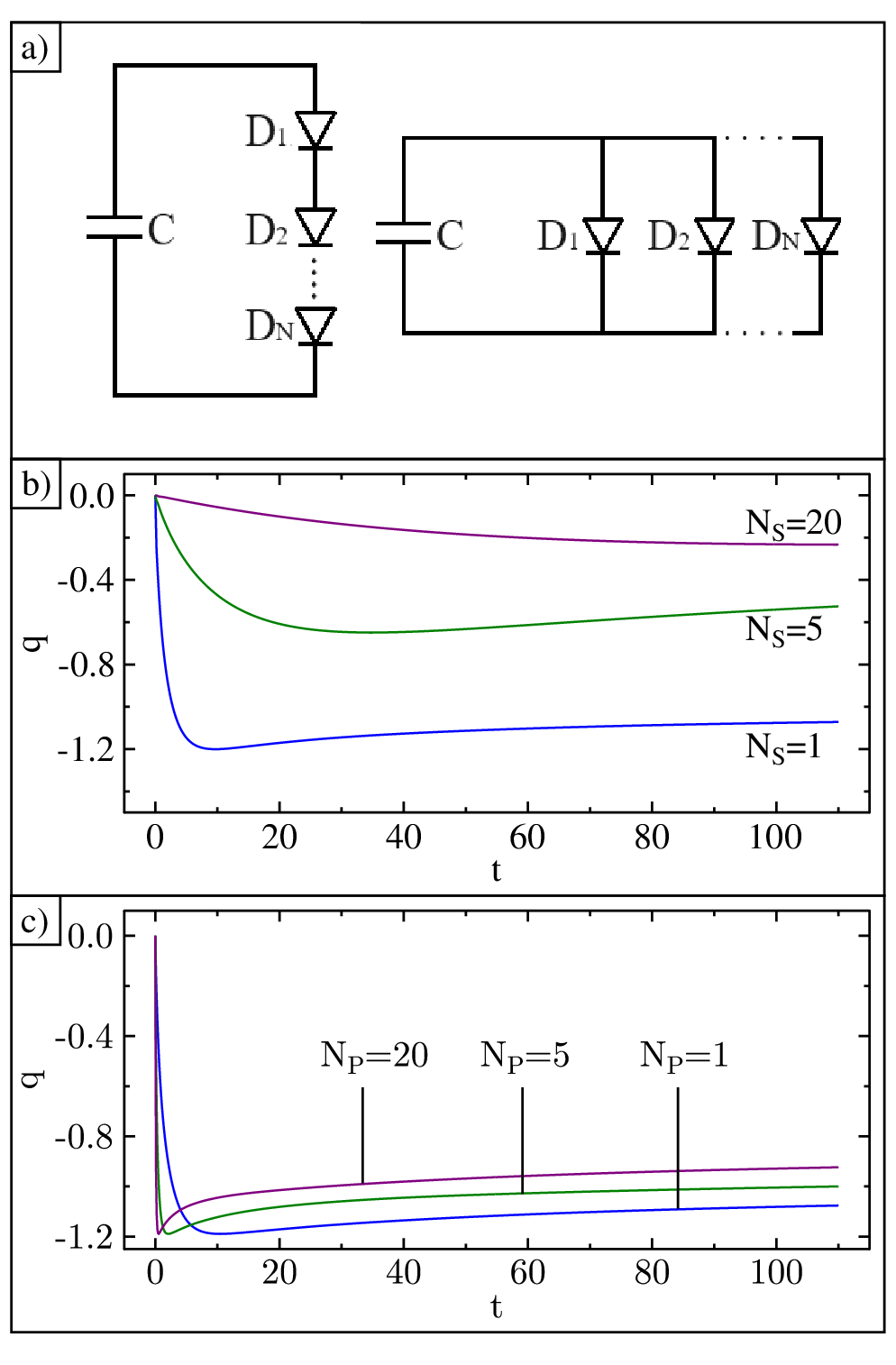}
\end{center}
\caption{Multiple diodes in series or parallel with a capacitor.~Numerical results use the following parameters: $u_0=0.025$, $R=1$, $C=4$, $k_BT=1$, and $V=0$.  (a)~Schematic of the series and parallel circuits. Average charge on the capacitor in time is shown for various numbers of diodes connected in (b)~series and (c)~parallel.}
\end{figure}

\begin{figure}[ht]
\begin{center}
\includegraphics[width=8cm]{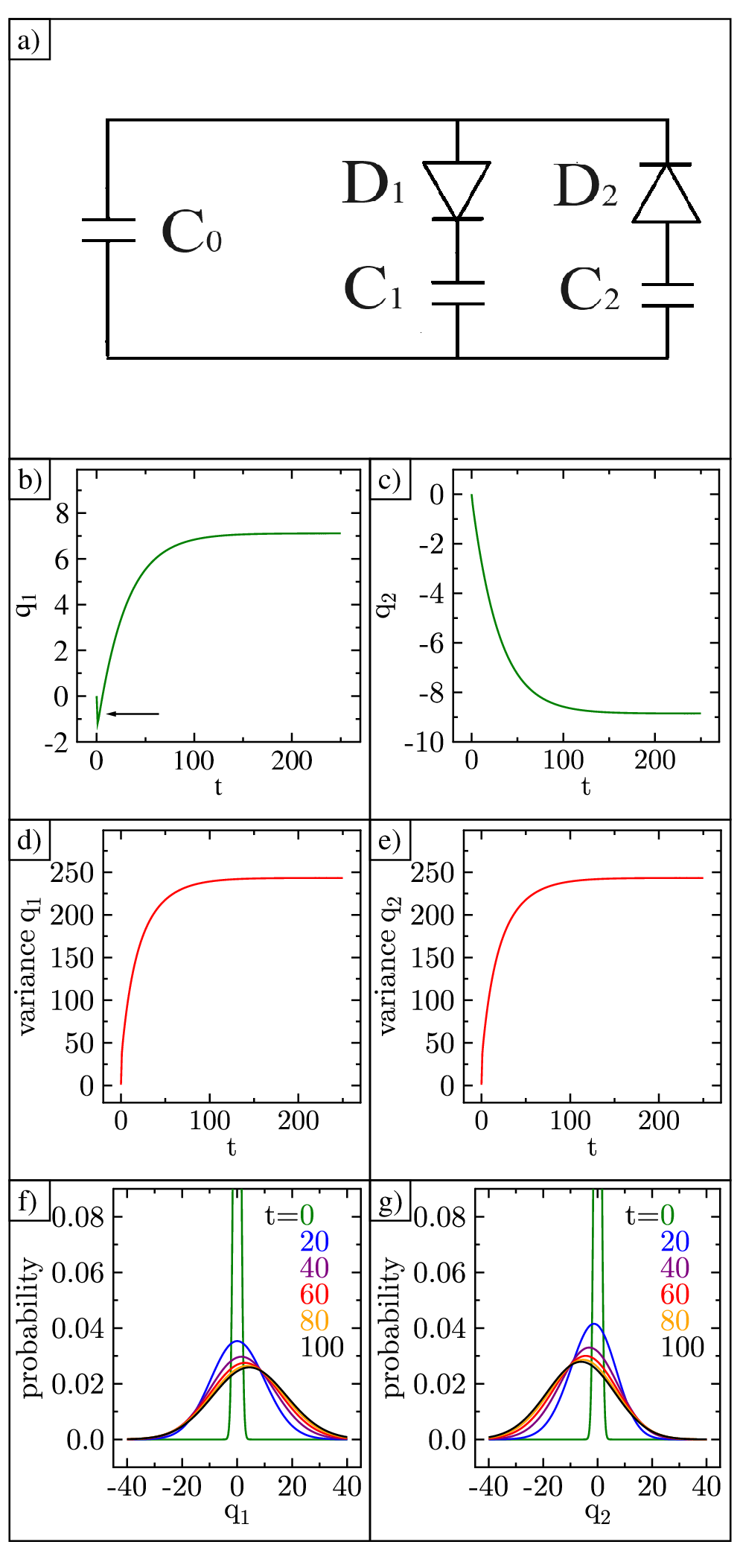}
\end{center}
\caption{Two-loop circuit dynamics using the following parameters: $k_BT_1=10$, $k_BT_2=1$, $u_0=1$, $R=0.1$, $C_0=4$, and $C_1=C_2=100$. (a)~Circuit schematic. Average charge in time on capacitor (b)~$C_1$ and (c)~$C_2$. Variance of the charge distribution in time on capacitor (d)~$C_1$ and (e)~$C_2$. Time evolution of charge probability distribution on capacitor (f)~$C_1$ and (g)~$C_2$.}
\end{figure}

The equivalent conductance for multiple identical diodes connected together in parallel is given by

\begin{eqnarray}
\mu_p(u)=\frac{N_p}{R} \frac{1}{1+e^{-u/u_0}}.
\end{eqnarray}
where $N_p$ is the number of diodes in parallel. In this case, the equivalent resistance decreases proportional to the number of diodes to become $R/N_p$. More importantly, notice the voltage drop across the diode is unchanged when compared to the single diode case. The average charge of the capacitor in time for different numbers of diodes in parallel is shown in Fig. 2(c). The maximum charge reached during charging is the same for each case. However, the time to reach the maximum charge decreases as the number of diodes increases. In this case, the lower overall resistance is responsible for the observed increase in the charging dynamics.

\maketitle
\begin{center}
\textbf{III. DOUBLE LOOP CIRCUIT}
\end {center}

The second circuit we studied is shown in Fig. 3(a). It has a three capacitors, two diodes, and two current loops. A complete time-dependent and time-independent theoretical analysis of this circuit occurs in the first paper of this series. Notice the circuit is similar to a full-wave rectifying circuit~\cite{har20}. The Hamiltonian for this circuit is given by

\begin{eqnarray}
\mathcal{H}=\frac{(q_1+q_2)^2}{2C_0}+\frac{q_1^2}{2C_1}+\frac{q_2^2}{2C_2},
\end{eqnarray}
where $q_1$ is the charge on capacitor $C_1$, $q_2$ is the charge on capacitor $C_2$, and the charge on capacitor $C_0$ can be written as $q_1+q_2$ using Kirchhoff's junction law. The two-loop circuit Fokker-Planck equation is given by

\begin{eqnarray}
\frac{\partial \rho}{\partial t}=\sum_{i=1}^{2}\frac{\partial}{\partial q_i}\left(\mu_i\rho\frac{\partial \mathcal{H}}{\partial q_i}\right)+k_BT_i\frac{\partial}{\partial q_i}\left(\mu_i\frac{\partial \rho}{\partial q_i}\right ),
\end{eqnarray}
where $\rho$ is now a function of $q_1$, $q_2$, and $t$. The diodes are wired oppositely but otherwise have identical conductance $\mu_i$ given by Eq. (2). The diode voltages $u_i$ are found using Eq. (6) and $T_i$ sets the temperature of each diode separately. 

\begin{figure}[ht]
\begin{center}
\includegraphics[width=8cm]{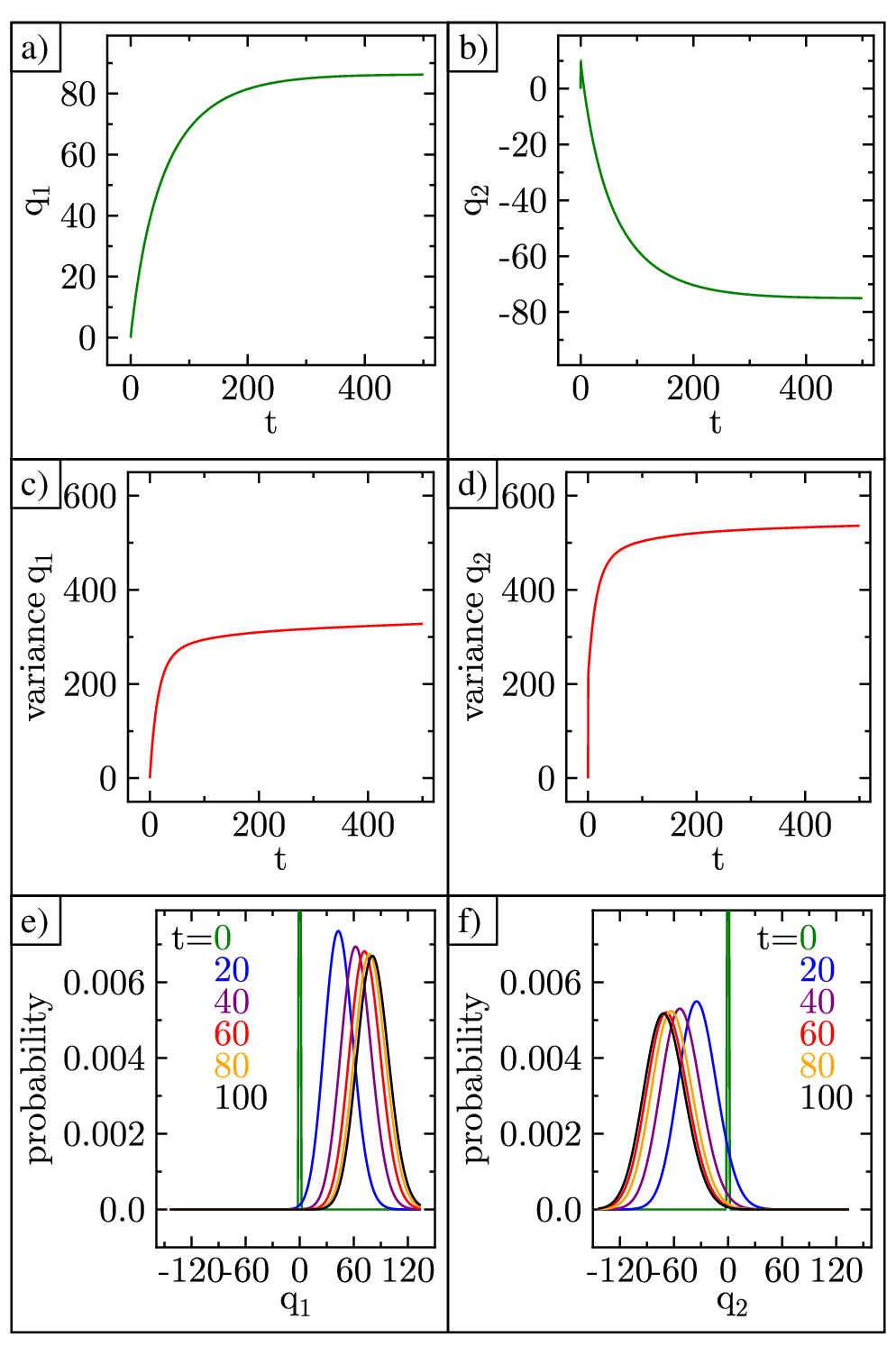}
\end{center}
\caption{Two-loop circuit dynamics using the following parameters: $k_BT_1=1$, $k_BT_2=100$, $u_0=1$, $R=0.1$, $C_0=4$, and $C_1=C_2=100$. Average charge in time on capacitor (a)~$C_1$ and (b)~$C_2$. Variance of the charge distribution on capacitor (c)~$C_1$ and (d)~$C_2$. Time evolution of the charge probability distribution on capacitor (e)~$C_1$ and (f)~$C_2$.}
\end{figure}

\maketitle
\begin{center}
\textbf{A. Transient results}
\end {center}

Numerical results for Eq. (7) using parameter values $k_BT_1=10$, $k_BT_2=1$, $R=0.1$, $u_0=1$, $C_0=4$, and $C_1=C_2=100$ are shown in Fig.~3. The charge on $C_1$ decreases for a very brief instant but then increases to a large positive value and stays constant as shown in Fig.~3(b). The charge on $C_2$ monotonically decreases to a large negative value and stays constant as shown in Fig. 3(c). Notice we are using a significantly larger value for the diode parameter for this circuit. In the first circuit, a diode parameter value this high resulted in almost no charging [see Fig. 1(f)], whereas here the charge is nearly 100 times larger. The magnitude of the final charge on the lower temperature branch capacitor ($C_2$) is slightly larger than the final charge on the higher temperature branch capacitor ($C_1$). Also, the initial charging spike and charge reversal, identified with the arrow in Fig. 3(b), only occurs on the higher temperature branch capacitor. The charge on $C_0$ is not shown but is given by $q_1+q_2$ and is therefore much smaller in magnitude than both $q_1$ and $q_2$. Having the two storage capacitors adjacent to the diodes drives the charging enhancement as charge can circulate between the diodes and charge the capacitors.

The variance of the charge distribution in time on capacitors $C_1$ and $C_2$ are shown in Figs. 3(d) and 3(e), respectively. In both cases, the variance, which is initially zero, quickly rises and then plateaus. The final variances are very similar even though the diode temperatures are ten times different. The time evolution of the charge distribution on $C_1$ is shown in Fig. 3(f). At time $t=0$ the plot closely resembles the delta function, with near zero average charge and variance. In time the peak of the distribution stays near zero at first, but then shifts to positive values. This tracks well with the average charge shown above. The time evolution of the charge distribution on $C_2$ is shown in Fig. 3(g). In time its peak moves to negative values. The final distributions on both capacitors are not Gaussian.

Additional numerical results for Eq. (7) are shown in Fig.~4. Here we set $k_BT_1=1$ and $k_BT_2=100$, but otherwise use identical parameters as above. Notice, for these numerical studies we have $k_BT_2$ larger than $k_BT_1$ to observe its role on the initial charging and on the sign of the charge. This time the charge on $C_1$ monotonically increases to a large positive value and stays constant as shown in Fig.~4(a). The sign is the same as before, but no initial negative charging occurs. The charge on $C_2$ increases for a very brief instant and then decreases to a large negative value and stays constant as shown in Fig. 4(b). Like before, the magnitude of the final charge on the lower temperature branch capacitor ($C_1$) is larger than the higher temperature branch capacitor ($C_2$). In both numerical studies, the magnitude of the initial brief charge is approximately equal to the difference in the two final charge magnitudes.

\begin{figure}[ht]
\begin{center}
\includegraphics[width=8cm]{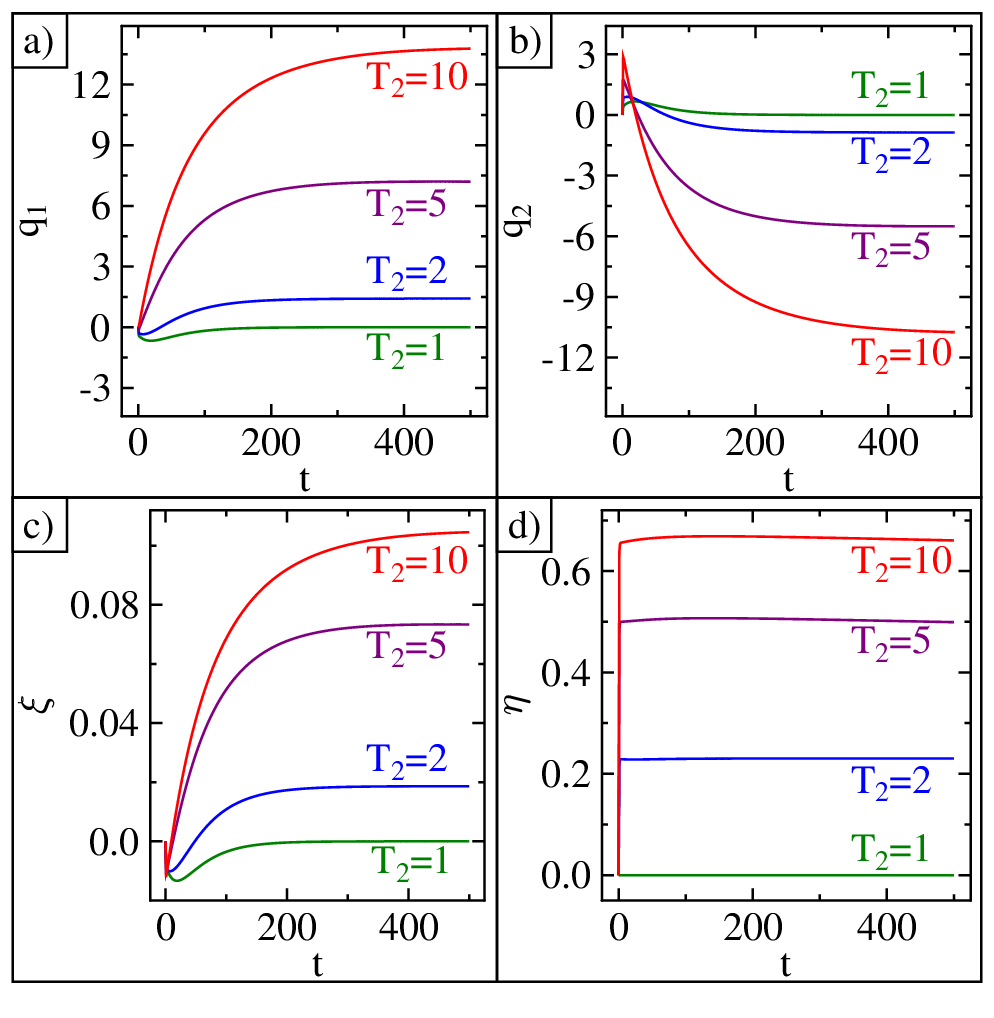}
\end{center}
\caption{Numerical results for the time-dependent Fokker Planck equation while varying $k_BT_2$. Parameter values are $u_0 = 0.2$, $R=0.1$, $C_0=4$, $C_1=C_2=100$, and $k_BT_1=1$. Average charge in time on capacitor (a)~$C_1$, and (b)~$C_2$. Average sum and difference scaled charges in time (c)~$\xi$ and (d) $\eta$.}
\end{figure}

The magnitude of the final charges in Fig.~4 are ten times larger than in Fig.~3. The sign of the final charge on $q_1$ is always positive and $q_2$ is always negative, and independent of which branch is at a higher temperature. The sign of the charge follows the high conductance direction of the diodes. In fact, a counterclockwise positive current flow is circulating between the two diode branches.

The charge variances on $C_1$ and $C_2$ are shown in Figs.~4(c) and 4(d), respectively. This time the variance on capacitor $C_2$ (the higher temperature branch) is noticeably larger. The time evolution of the probability distributions on $C_1$ and $C_2$ are shown in Figs. 4(e) and 4(f), respectively. In time they evolve in the same directions as earlier, but the shift is larger. It is also easy to see that the peaks are larger and the widths narrower for $C_1$ (the lower temperature branch).

\begin{figure}[ht]
\begin{center}
\includegraphics[width=8cm]{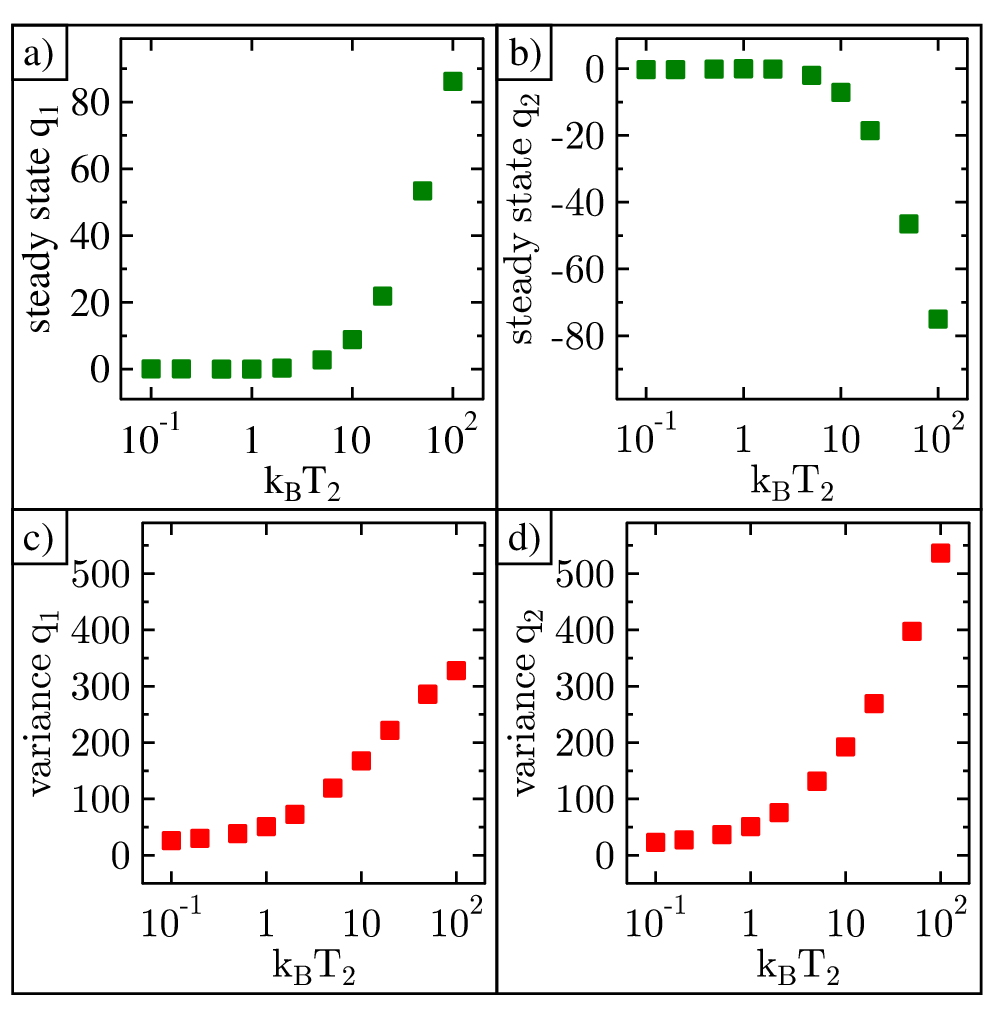}
\end{center}
\caption{Numerical results using parameter values of $k_BT_1=1$, $u_0=1$, $R=0.1$, $C_0=4$, and $C_1=C_2=100$. Steady state charge vs $k_BT_2$ on capacitor (a)~$C_1$ and (b)~$C_2$. Variance of charge on capacitor (c)~$C_1$ and (d)~$C_2$.}
\end{figure}

Next we ran additional numerical studies for Eq. (7) using a lower value of $u_0$ to slow down the initial charging dynamics. Using $u_0=0.2,$ $k_BT_1=1,$ and varying $k_BT_2$ the average charge on  $C_1$ and $C_2$ are shown in Figs. 5(a) and 5(b), respectively. Notice the height of the initial spike increases with temperature but the time duration of the initial spike is longer as the two diode temperatures become closer to equal as shown in Fig.~5(b).

In order to more easily compare our numerical results with the analytical results found in the first paper of this series, we plot the average charge data in terms of the sum and difference dimensionless charge variables given by the following 

\begin{eqnarray}
\xi=\epsilon\frac{q_1-q_2}{C_0V_0}
\end{eqnarray}

\begin{eqnarray}
\eta=\frac{q_1+q_2}{C_0V_0}(1+\epsilon),
\end{eqnarray}
where $\epsilon=C_0/2C$ and $V_0=\sqrt{k_B(T_1+T_2)/2C_0}$.

Notice the difference in charges is given by $\xi$, while the sum is given by $\eta$. The difference of the average charges shows the initial negative spike and then reverses sign as shown in Fig. 5(c). Also, notice the initial spike widens in time as the two temperatures become closer to equal. The sum of the average charges quickly increases to its maximum value and then plateaus as shown in Fig. 5(d). The sum reflects the charge on the smaller $C_0$ capacitor, and its dynamics are extremely quick. These results agree well with the analytical solutions discussed in the first paper of this series.

\begin{figure}[ht]
\begin{center}
\includegraphics[width=8cm]{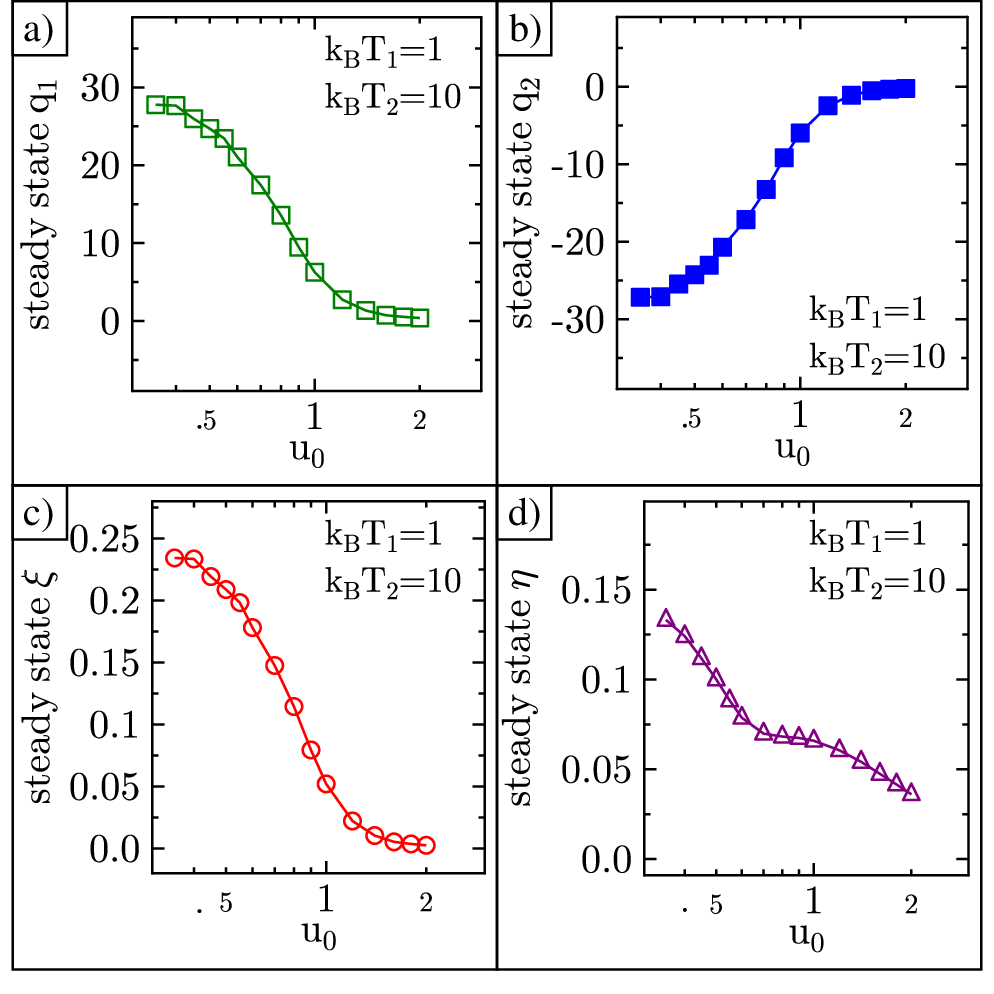}
\end{center}
\caption{Numerical results for the time-independent Fokker Planck equation. Fixed parameter values are $R=0.1$, $C_0=4$, $C_1=C_2=100$, $k_BT_1=1$, and $k_BT_2=10$. The steady charge on capacitor (a)~$C_1$, (b)~$C_2$, and (c)~$C_0$. (d) The difference in capacitor $C_1$ and $C_2$ charge.}
\end{figure}

\maketitle
\begin{center}
\textbf{B. Steady state charge and variance}
\end {center}

Very long time simulations of Eq. (7) using a broad range of $k_BT_2$ values were completed and analyzed. The other parameter values were set to be $R=0.1$, $u_0=1$, $C_0=4$, $C_1=C_2=100$, and $k_BT_1=1$. Generally, after a long time period, each capacitor charge stopped changing. We assume the value of this final charge is the steady state solution to Eq. (7). The final charges on $C_1$ and $C_2$ vs $k_BT_2$ are shown in Fig.~6(a) and 6(b), respectively. As before, the charge is always positive on $C_1$ and negative on $C_2$. Also, the magnitude of charge on the lower temperature branch capacitor $C_1$ is always slightly larger in magnitude. The charge variances, at the same final time, on capacitors $C_1$ and $C_2$ are shown in Fig.~6(c) and 6(d), respectively. The variance on both capacitors grows as $k_BT_2$ increases, even though $k_BT_1$ is fixed.

\
\
\

\vskip 12 pt
\vskip 12 pt
\vskip 12 pt

\maketitle
\begin{center}
\textbf{C. Time-independent numerical studies}
\end {center}

In order to more easily compare our numerical results with the analytical results, we also solved for the stationary probability density within the Chapman-Enskog derivation found in the first paper of this series. One benefit of these solutions is that we can significantly expand our parameter range for $u_0$. Essentially, we found the time-independent solution to Eq.~(7) in integral form in the first paper of this series. Using the Chapman-Enskog method, the equation was scaled and solved for the parameters given by eqns. (8) and (9)~\cite{chapman, enskog, bon14, bon19, ris84, gardiner}. The steady state average charges $q_1$ and $q_2$ for different values of $u_0$ are shown in Figs.~7(a) and 7(b). As $u_0$ decreases, both steady state charge magnitudes increase by another factor of 10 before plateauing at $u_0=0.4$. As with the time-dependent solutions, the capacitor on the cooler branch, $q_1$ has a larger charge. Plots for dimensionless sum and difference charges $\xi$ and $\eta$ are shown in Figs.~7(c) and 7(d). Both parameters increase in magnitude as $u_0$ decreases. All of these results are in excellent agreement with the analytical results found in the first paper of this series.

\
\maketitle
\begin{center}
\textbf{V. CONCLUSIONS}
\end {center}

In this study, we present coordinated results for the role nonlinear devices play in thermal energy harvesting across two papers in a series. The first paper presents analytical solutions. In this second paper, we present numerical solutions.

We simulated the charge dynamics for two circuits by solving the time-dependent Fokker Planck equation. The first circuit has a single loop connecting a DC bias voltage, a capacitor, and a diode. The capacitor charges during the transient phase in all cases. The maximum charge reached increases with increasing temperature and capacitance. In addition, as the diode performs more like an ideal switch, the maximum charge increases and the time to discharge increases. We also showed that multiple diodes in parallel lowers the circuit resistance and allows the maximum charge to be reached more quickly, while multiple diodes in series significantly reduces the maximum charge.

The second circuit we analyzed has two loops, one small capacitor, two storage capacitors, and two diodes. We found the charge on the two storage capacitors increases far above that reached by the single loop circuit. One capacitor has a large positive charge and the other has a large negative charge due to how the diodes are wired in opposition. The capacitor on the lower temperature branch ends up with a slightly larger charge and a slightly smaller variance, as expected.

For the two-loop circuit we also solved the time-independent Fokker Planck equation. We determined the steady state capacitor charges across a broad range of diode parameter values $u_0$. As $u_0$ tends toward zero the diode becomes an ideal switch and the circuit builds up the most charge and holds its charge longer. As $u_0$ becomes larger the diode quality drops until it becomes a simple linear resistor. When the diodes are replaced with resistors, the circuits cannot accumulate any charge at any time or at any temperature difference. Diode nonlinearity and orientation are key drivers for capacitor charging from the thermal environment.

The next natural step for this study is to perform experimental tests of these predictions. The circuit shown in Fig.~3(a) will be tested first. One idea is to place $D_1$ in liquid nitrogen and leave the other components at room temperature. Based on the numerical studies presented here we will make the capacitance of $C_0$ small and make the other two large. We expect the two storage capacitors to build up a permanent charge and that the charges will be equal but opposite in sign.


\maketitle
\begin{center}
\textbf{VI. ACKNOWLEDGMENTS}
\end {center}

This work was financially supported, in part, by an award from the WoodNext Foundation (award number AWD-106363), which is administered by the Fidelity Investments Charitable Gift Fund, and by the FEDER/Ministerio de Ciencia, Innovaci\'on y Universidades -- Agencia Estatal de Investigaci\'on  (MCIN/AEI/10.13039/501100011033) grants PID2020-112796RB-C22 and PID2024-155528RB-C22.

\end{document}